\documentclass[prl,
               twocolumn,
                preprintnumbers,
                amsmath,
               amssymb,
                showpacs,
               superscriptaddress]{revtex4}
\usepackage{graphicx} 
\usepackage{dcolumn}  
\usepackage{bm}       
\usepackage{amsfonts}
\usepackage{dsfont}
\usepackage{xcolor}

\usepackage{booktabs}

\usepackage{ulem}
\usepackage{color}

\begin{document}

\title{Angle-Resolved Attosecond Streaking of Twisted Attosecond Pulses}

\author{Irfana N. Ansari}
\affiliation{%
Department of Physics, Indian Institute of Technology Bombay,
            Powai, Mumbai 400076, India}

\author{Deependra S. Jadoun}
\affiliation{%
Department of Physics, Indian Institute of Technology Bombay,
            Powai, Mumbai 400076, India}

\author{Gopal Dixit}
\email[]{gdixit@phy.iitb.ac.in}
\affiliation{%
Department of Physics, Indian Institute of Technology Bombay,
            Powai, Mumbai 400076, India}

\date{\today}


\begin{abstract} 
The present work focuses on the characterisation  of the amount of orbital angular momentum
(OAM) encoded in  the twisted attosecond pulses via energy- and angle-resolved attosecond 
streaking in pump-probe setup. It is found that the photoelectron spectra generated by the 
linearly polarised twisted pulse with different OAM values exhibit angular modulations, whereas 
circularly   polarised twisted pulse yields angular isotropic spectra. It is demonstrated that the 
energy- and angle-resolved  streaking spectra are sensitive to the OAM values of the twisted 
pulse. Moreover, the different combinations of the polarisation of the twisted pump pulse and 
strong infrared probe pulse influence the streaking spectra differently.   The 
characterisation of the OAM carrying twisted attosecond pulses opens up 
the possibility to explore helical light-matter interaction on attosecond timescale. 
\end{abstract}

\maketitle 

Complete characterisation of attosecond pulses is essential to harness the full potential of such 
pulses in capturing ultrafast electron processes  and extracting meaningful interpretations of 
time-resolved measurements~\cite{krausz2014attosecond, krausz2009attosecond}. 
Different techniques are in practice to 
characterise the temporal structure of the 
attosecond pulses such as 
attosecond streaking and RABBITT~\cite{itatani2002attosecond, kitzler2002quantum, paul2001observation}.
Both these methods yield information about pulse duration, carrier energy,  
and chirp of the attosecond pulses~\cite{mairesse2005frequency, quere2005temporal}.
Recently, attosecond streaking method has been extended to 
characterise the  carrier-envelope phase~\cite{he2016carrier} and 
extract the polarisation state~\cite{jimenez2018attosecond} of the isolated attosecond  pulse. 
In this work, energy- and angle-resolved  attosecond streaking (AAS) method is employed to 
characterise the orbital angular momentum (OAM) of twisted attosecond pulses.

The polarisation property of light is associated with its spin angular momentum, 
whereas  the spatial profile of the wave front is connected to the OAM of the light~\cite{allen1992orbital}.  
Light with non-zero OAM is known as twisted light
and Laguerre-Gaussian light beam is an example of such light and considered in this work. 
OAM of the light has found numerous applications in various fields since its first realisation~\cite{cardano2015spin, babiker2018atoms, furhapter2005spiral, forbes2019raman, brullot2016resolving, forbes2018optical, andersen2006quantized, inoue2006entanglement, ruchon2020magnetic}.  
These applications have motivated scientists to generate ultrashort twisted pulses in the extreme ultraviolet (XUV)  energy regime to probe attosecond electron motion. 
In recent years, series of works have been carried out to 
up-convert OAM carrying infrared (IR) beam to the XUV beam 
using high harmonic generation (HHG)~\cite{zurch2012strong, hernandez2013attosecond, gariepy2014creating, geneaux2016synthesis, rego2016nonperturbative, turpin2017extreme, hernandez2017extreme, paufler2018tailored, gauthier2019orbital}. 
A  non-collinear method  has been applied to generate linearly polarised twisted 
light with relatively low OAM via  HHG~\cite{gauthier2017tunable, kong2017controlling, 
dorney2019controlling}.  
These state-of-the-art HHG experiments provided an avenue for the generation of 
coherent attosecond XUV pulses with  desirable OAM properties. 
However, no claim of generating such pulses can be made without  
measurement of the OAM encoded in the twisted XUV pulses 
with no {\it a priori} assumptions. This  is an uncharted territory and  
main focus of the present work.  

In this work, the concept of the standard attosecond streak camera is extended in
three-dimensions as the twisted XUV pulse has a complex spatial structure. 
The energy- and angle-resolved photoelectron spectra are simulated as a function of pump-probe delay time for 
different polarisations of the twisted XUV pump pulse and IR probe pulse. The 
pump pulse induces a single-photon ionisation and liberated photoelectron is streaked 
by the synchronized  IR pulse having the plane wavefront and  zero OAM. 
Kaneyasu {\it et al.} have performed the photoionisation of helium by the twisted 
XUV beam at synchrotron and discussed the possibility of observing 
the violation of the standard electric dipole selection rules~\cite{kaneyasu2017limitations}.  
Boning {\it et al.}  have discussed the streaking of twisted x-waves 
within dipole approximation 
with relatively weak continuous IR beam~\cite{boning2017attosecond}.  
The influences of the projection of the total angular momentum, the opening angle and the impact parameter on the streaking spectra have been discussed~\cite{boning2017attosecond}. Recently, the formalism of two-photon ionisation and its consequences in 
photoionisation time delay have been discussed~\cite{giri2020signatures}.

The transition amplitude, within strong-field approximation,
from a ground state to a continuum state $|k \rangle$ with momentum $\mathbf{k}$ is expressed as 
\begin{equation}\label{eq01}
a_{\mathbf{k}} (\tau)  =  -i \int_{-\infty}^{+\infty} dt  e^{ i [\textrm{I}_\textrm{p} t - \Phi(\mathbf{k}, t)]} 
\mathbf{E}_{\textrm{X}}(t-\tau) \mathbf{d}_{\mathbf{p}(t)},
\end{equation}
where $\textrm{I}_\textrm{p}$ is the ionisation potential of the atom, 
$\Phi(\mathbf{k}, t) = \int_{t}^{+\infty} dt^{\prime} \mathbf{p}^{2}(t^{\prime})/2$ is Volkov phase 
with $\mathbf{p}(t) = \mathbf{k}+\mathbf{A}(t)$ as the instantaneous momentum of the 
photoelectron in IR field, $\mathbf{A}(t)$ being the vector potential 
such that $\mathbf{E}(t) = -\partial \mathbf{A}(t)/\partial t$. $\mathbf{E}_{\textrm{X}}(t) = 
\mathbf{\tilde{E}}_{\textrm{X}}(t) \textrm{exp}[-i \Omega t]$ is the XUV field with 
$\mathbf{\tilde{E}}_{\textrm{X}}(t)$ as the envelope and $\Omega$ as the central 
energy of the XUV field. The parameter $\tau$ is the time delay between the two pulses, and 
$\mathbf{d}_{\mathbf{p}(t)}$ is the transition amplitude  from the ground state to the 
continuum state $\vert \textbf{p}(t)\rangle$ with kinetic momentum $\textbf{p}(t)$. 
In the presence of an IR field, the phase accumulated by the photoelectron,  
during its motion in the continuum from $t$ to $+\infty$, is 
$\delta \Phi(t) = \Phi(\mathbf{k}, t) - \textrm{I}_\textrm{p} t$. Note that strong-field approximation
 is appropriate as  $\omega \ll \textrm{I}_\textrm{p}$ with $\omega$ 
 as the frequency of the IR field, which only interacts with the emitted photoelectron. 

Twisted XUV pulse induced  photoionisation and 
exchange of the OAM larger than one unit in photoionisation has been manifested. 
As  a result, the standard dipole selection rules get modify~\cite{picon2010transferring, picon2010photoionization}. 
The modified  selection rules for electronic transitions read as $|l_{f}-l_{i}| \leq \vert l \vert +1$,  $m_{f}-m_{i} = l \pm 1$ and $l_{f}-l_{i} + |l| +1$ is even. 
Here, $l_{f} (m_{f})$ and $l_{i} (m_{i})$ are the final and the initial orbital angular   
(magnetic) quantum numbers of electronic states, respectively; 
and $l$ is the topological charge of the  twisted XUV pulse. 
It is evident from the selection rules that  only two values of $l_{f} = l \pm 1$ are allowed 
when the twisted XUV pulse with $l \geq 0$ ionises hydrogen atom ($l_{i} = 0$). 
In such situation, photoionisation transition  amplitude reads as
\begin{eqnarray}\label{eq02}
\mathbf{d}_{\mathbf{k}} & = & \frac{i \Omega \sqrt{2^{l}}}{2 \pi w_{0}^{l+1}} \left[ \frac{C_{1}i^{l+1}}{(-1)^{l+1}} \ d_{k}^{l+1} \textit{Y}_{l+1}^{l+1}(\theta_{k},\phi_{k})   \right.\nonumber \\
&& +  \frac{C_{2}}{(-1)^{l}} 
\left\{ a\ i^{l-1}\ d_{k}^{l-1} \textit{Y}_{l-1}^{l-1}(\theta_{k},\phi_{k}) 
\right. \nonumber \\
&& \left. \left. \hspace{5em} +b\ i^{l+1}\ d_{k}^{l+1}  \textit{Y}_{l+1}^{l-1}(\theta_{k}, \phi_{k})  \right\} \right]. 
\end{eqnarray}
Here, $w_{0}$ is the beam waist; and $C_{1} = 2^{l}(l + 1)! \ [4\pi/(2l+3)!]^{1/2}$, 
$C_{2} = 2^{l+1}\ l!\ \pi [2/ 3(2l+1)!]^{1/2}$, $a = [3(2l+1)/4\pi(2l-1)]^{1/2}\ C_{l,1,0,0}^{l-1,0}\ C_{l,1,l,-1}^{l-1,l-1}$ and $b = [3(2l+1)/4\pi(2l+3)]^{1/2}\ C_{l,1,0,0}^{l+1,0}\ C_{l,1,l,-1}^{l+1,l-1}$ are constants. 
The Clebsch-Gordon coefficients $C_{l_{1},l_{2},m_{1},m_{2}}^{l_{f},m_{f}}$ are used to 
express  constants $a$ and $b$. To obtain the above equation, hydrogen ground state wave 
function is written as a product of the radial wave  function and the spherical harmonic 
$\textit{Y}_{l_{i}}^{m_{i}}$, whereas the wave function for the continuum state is expanded in terms of the spherical Bessel function of first kind $j_{l_{f}}$ and the spherical harmonics. 
The radial transition amplitude is $d_{k}^{l} = \int_{0}^{\infty} dr\ r^{l +3}~j_{l_{f}}(kr) \textit{R}_{n_{i}, l_{i}}(r)$ where $ \textit{R}_{n_{i}, l_{i}}$ is the radial ground-state  wave function. Eq.~(\ref{eq02}) is obtained for twisted XUV pulse, which is linearly  polarised along $x$-axis and  propagating along $z$-axis. The form of the vector potential for the Laguerre-Gaussian pulse is taken from Ref.~\cite{picon2010transferring, picon2010photoionization}.

Due to non-zero OAM of the twisted XUV pulse, three ionising paths are allowed  
from the unpolarised ground state of hydrogen as evident from Eq.~(\ref{eq02}). The strength 
of these paths depend on the magnitudes of $d_{k}^{l}$, $\textit{Y}_{l}^{m}$ and the 
constants. Possibility of more than one ionising path is a consequence of the modified 
selection rules.  As evident from Eq.~(\ref{eq02}), the three paths have different contributions from the spherical harmonics $\textit{Y}_{l}^{m}$, which depends on the OAM values and determine the resultant
angular distribution of the photoelectrons. 
Let us analyse how the photoelectron spectra are sensitive to the OAM of the ionising  XUV pulse. 
 
\begin{figure}[]
\includegraphics[width=8.5 cm]{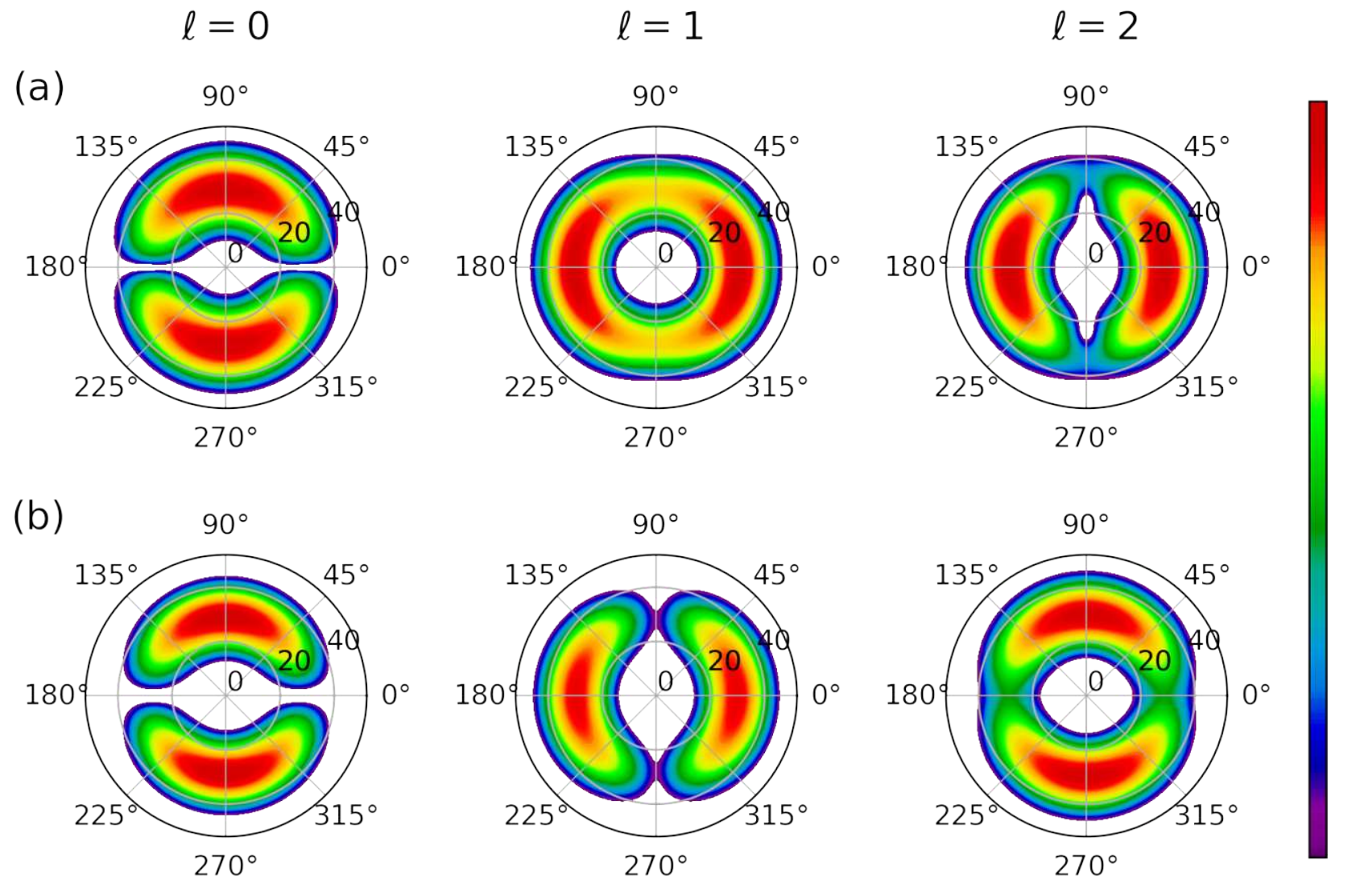}
\caption{Normalised photoelectron spectra, induced  by twisted XUV pulse, 
for hydrogen. The spectra are shown as a function of the observational azimuthal  angle $\phi_{k}$, the kinetic energy of photoelectron $E_{k}$ (shown radially, in eV) and the OAM values of the ionising twisted pulse with Gaussian envelop of 150 attoseconds pulse duration and 46.6 eV photon energy. 
The spectra  are shown for the observational polar  angle (a) $\theta_{k} = 30^{\circ}$ and (b) $\theta_{k} = 90^{\circ}$ (shown row-wise).}
\label{fig1}
\end{figure}

\begin{figure}[b!]
\includegraphics[width=8.5 cm]{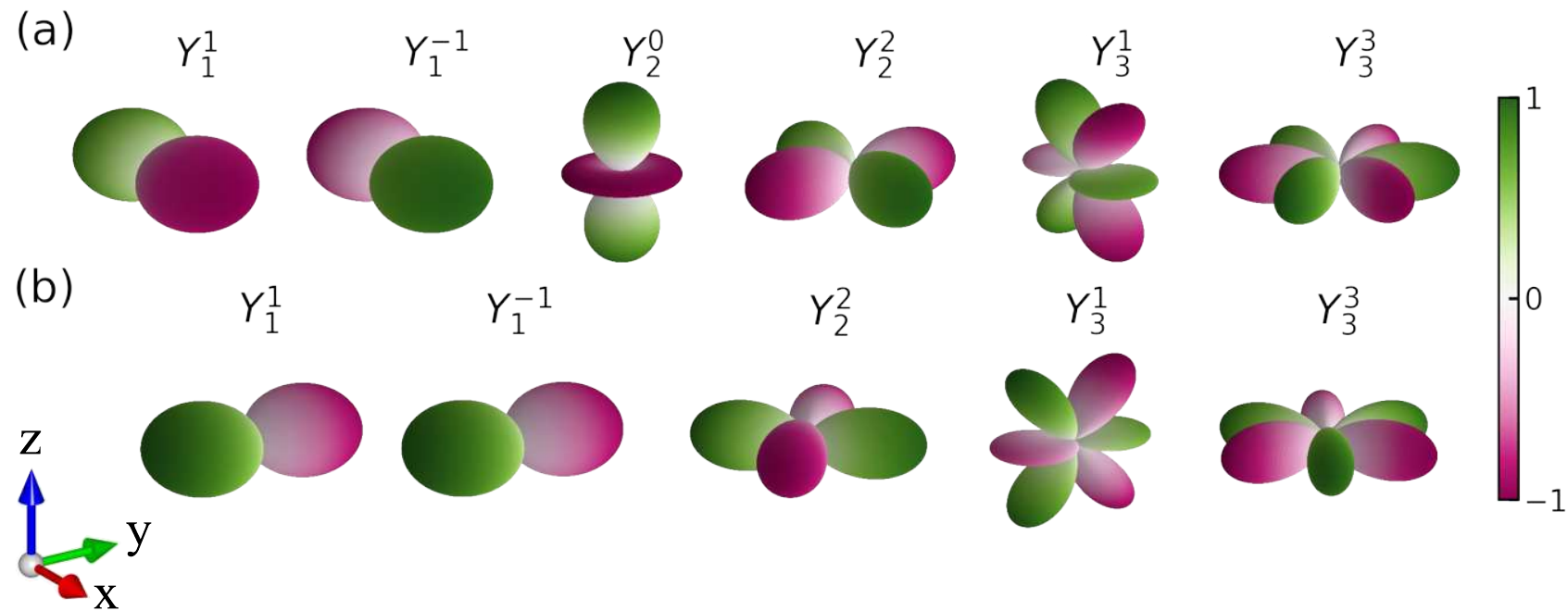}
\caption{Normalised (a) real and (b) imaginary parts of the relevant spherical harmonics. The spherical harmonic $Y_{2}^{0}$ is real and Im($Y_{2}^{0}$) is not shown.}
\label{fig01}
\end{figure}

The photoionisation (or unstreaked) spectra, induced by twisted pulse,  for different values of  the OAM are presented in Fig.~\ref{fig1}. The spectra are shown as 
a function of the observation azimuthal  angle $\phi_{k}$ and the kinetic energy of the 
photoelectron $E_{k}$ (shown radially in eV).  
The  spectra for  two observational polar angles $\theta_{k} = 30^{\circ}$ and $
\theta_{k} = 90^{\circ}$ are shown in Figs.~\ref{fig1}(a) and \ref{fig1}(b), respectively. The first 
and third ionisation paths contribute for $l = 0$ and their photoelectron 
distributions are given by $Y_{1}^{1}$ and $Y_{1}^{-1}$, respectively [see Eq.~(\ref{eq02})]. 
The nodes in the angular distribution along $x$-axis  at $\phi_{k} = 0^{\circ}$ and $\phi_{k} = 180^{\circ}$  [see Fig. \ref{fig01}(a) for $l = 0$] are generated due to the fact that the mentioned
paths contain equal and opposite real contributions  [see Fig. \ref{fig01}(a)].
In contrast to this, the second ionisation path is non-zero for  $l = 1$, but contributes 
a constant background to the total spectra as 
$Y_{0}^{0}$ is  angular isotropic. 
The other two contributing terms, $Y_{2}^{2}$ and $Y_{2}^{0}$ (Fig. \ref{fig01}), interfere 
destructively due to the pre-factors resulting in minimum intensities of the photoelectron  
distribution  along $y$-axis, i.e., at $\phi_{k} = 90^{\circ}$ and $\phi_{k} = 270^{\circ}$. 
The  qualitative behavior of the photoelectron angular distributions for different values of $\theta_{k}$ is  similar for $l = 0$ and 1 as visible from Fig. \ref{fig1}.  

In case of the twisted pulse with $l = 2$, all the three ionisation paths contribute differently for smaller and higher values of  $\theta_{k}$. 
The reason lies in the $\theta_{k}$-dependent distributions of the relevant spherical harmonics, specifically $Y_{3}^{1}$. 
The first, second and third ionisation 
paths correspond to $Y_{3}^{3}$, $Y_{1}^{1}$ and $Y_{3}^{1}$, respectively. 
The real parts of $Y_{3}^{3}$ and $Y_{3}^{1}$ interfere destructively along $x$-axis whereas their imaginary parts interfere constructively along $y$-axis for $\theta_{k} = 30^{\circ}$. 
Also, $Y_{1}^{1}$ contributes only along $x$-axis and results in a maxima along this direction. 
Thus, in this case, the angular distribution is dominated by  the real and imaginary parts of $Y_{3}^{3}$. 
However, both the real as well as the imaginary parts of $Y_{3}^{3}$ and $Y_{3}^{1}$ 
interfere destructively when $\theta_{k}$ is close to $90^{\circ}$ (see Fig.~\ref{fig01}).  
Therefore, the minima in the angular distribution along $y$-axis disappears and appears along 
$x$-axis as $\theta_{k}$ increases from smaller to larger values. 

From the discussion of Eq.~(\ref{eq02}) and analysis of  Fig.~\ref{fig1}, 
it is established that the unstreaked photoelectron spectra are sensitive to the OAM value of the linearly-polarised ionising pulse. 
It appears that the presence of the OAM in the ionising pulse can be determined by locating the intensity minima in the unstreaked spectra for smaller values of $\theta_{k}$ (say $\theta_{k} = 30^{\circ}$). 
However, the amount of the OAM can not be extracted because of the similar qualitative behavior of the unstreaked spectra for $l=1$ and $l=2$ [Fig. \ref{fig1}(a)]. Additionally, the unstreaked spectra are not helpful in even determining the presence of the OAM for higher values of $\theta_{k}$ [Fig. \ref{fig1}(b)].
Moreover, if circularly  polarised twisted XUV pulse is used to ionise the atom, the resultant photoelectron angular distributions  are isotropic irrespective of  
the helicity of the ionising XUV pulse. 
The spectra  are similar for different OAM values  as well (not shown here).
Thus, it can be concluded that the unstreaked spectra can only determine the presence or absence of the OAM in the ionising  
pulse, but with many restrictions on the properties of the pulse and the experimental 
observational setup.

To characterise the exact amount of the OAM encoded in the twisted  pulse, 
it is important to introduce IR pulse to streak the liberated photoelectron. 
As  the unstreaked  spectra are sensitive to the observational angle, it is meaningful to  use circularly  polarised IR pulse, which is represented as  
 \begin{equation}\label{eq03}
\mathbf{E}_{L}(t)= \frac{\mathcal{E}_{L}(t)}{\sqrt{2}}\  \left[ \cos (\omega_{L}t)\hat{x} -\Lambda_{L} \sin (\omega_{L}t)\hat{y}\right],
\end{equation} 
where $\mathcal{E}_{L}(t)$ and $\Lambda_{L}$ are the envelop and helicity of the IR pulse, respectively.  In this case, the Volkov phase reads as
\begin{equation}\label{eq04}
\Phi(\mathbf{k},t)=-[k^{2}+\mathcal{A}_{L}^{2}(t)]\frac{t}{2}+\frac{k\mathcal{A}_{L}(t)}
{\omega_{L}}\sin\theta_{k}\cos(\omega_{L}t+\Lambda_{L}\phi_{k}).
\end{equation}
In the following, right-handed circularly polarised IR pulse with 800 nm wavelength is used to streak the photoelectrons. The pulse duration and intensity of the IR pulse are 5 
femtoseconds with Gaussian envelope and $5\times10^{13}$ W/cm$^{2}$, respectively.  
To get the streaking spectra, $\mathbf{d}_{\mathbf{k}}$ is replaced by 
 $\mathbf{d}_{\mathbf{p}(t)}$. 
 
\begin{figure}[t!]
\includegraphics[width=8.5 cm]{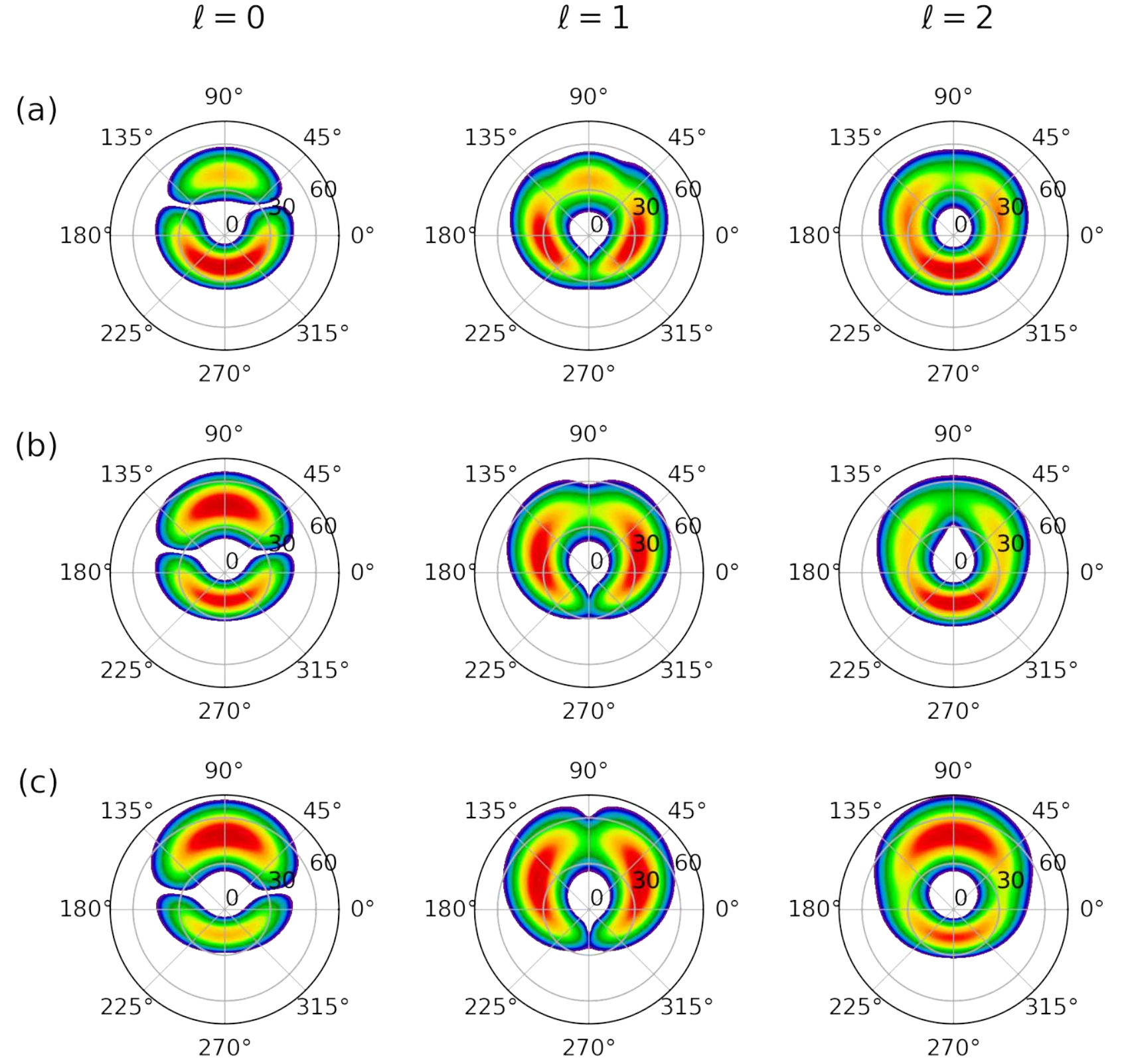}
\caption{AAS spectra for different OAM values. 
The twisted XUV pulse with linear polarisation is used to trigger  photoionisation 
and right-handed  circularly  polarised IR pulse is used to streak the liberated photoelectrons. The spectra are plotted as a functions of $\phi_{k}$, $E_{k}$ (shown radially, in eV) and 
for (a) $\theta_{k} = 30^{\circ}$, (b) $\theta_{k} = 45^{\circ}$ and (c) $\theta_{k} = 60^{\circ}$. All the spectra are normalised with respect to their respective maxima.} 
\label{fig2}
\end{figure}

AAS spectra for different OAM values are presented in Fig.~\ref{fig2}. 
The vector potential of the streaking IR pulse is directed along negative $y$-axis at the instant of ionisation when the time-delay between the twisted XUV and IR pulses is zero. 
This streaking field induces a change in momentum of the photoelectron $\delta \textbf{k}=-\textbf{A}_{L}(t)$~\cite{itatani2002attosecond}.
It results in the streaking of the photoelectrons along  $+y$-direction, i.e.,  $\phi_{k} = 
90^{\circ}$. At a first glance, it is evident that the streaking 
spectra are sensitive to the OAM values 
as well as to the observation direction of the photoelectrons. 
On a close inspection of the streaking spectra when $\theta_{k}=30^{\circ}$, 
it is clearly visible that the low-energy regime of the spectra around $\phi_{k} = 270^{\circ}$ are 
most intense for $l = 0$ and 2, whereas it is least intense around this region for $l = 1$ (Fig. \ref{fig2}).  Moreover, the numbers of intensity minima in the 
spectra are 2 for $l = 0$, and 3 for $l = 1$ and $l = 2$; 
and the angular positions of these minima are different in different spectra. 
Similar observations can be made for $\theta_{k} = 45^{\circ}$ and $60^{\circ}$ 
[Fig. \ref{fig2}(b, c)].

\begin{figure}[t!]
\includegraphics[width=8.5 cm]{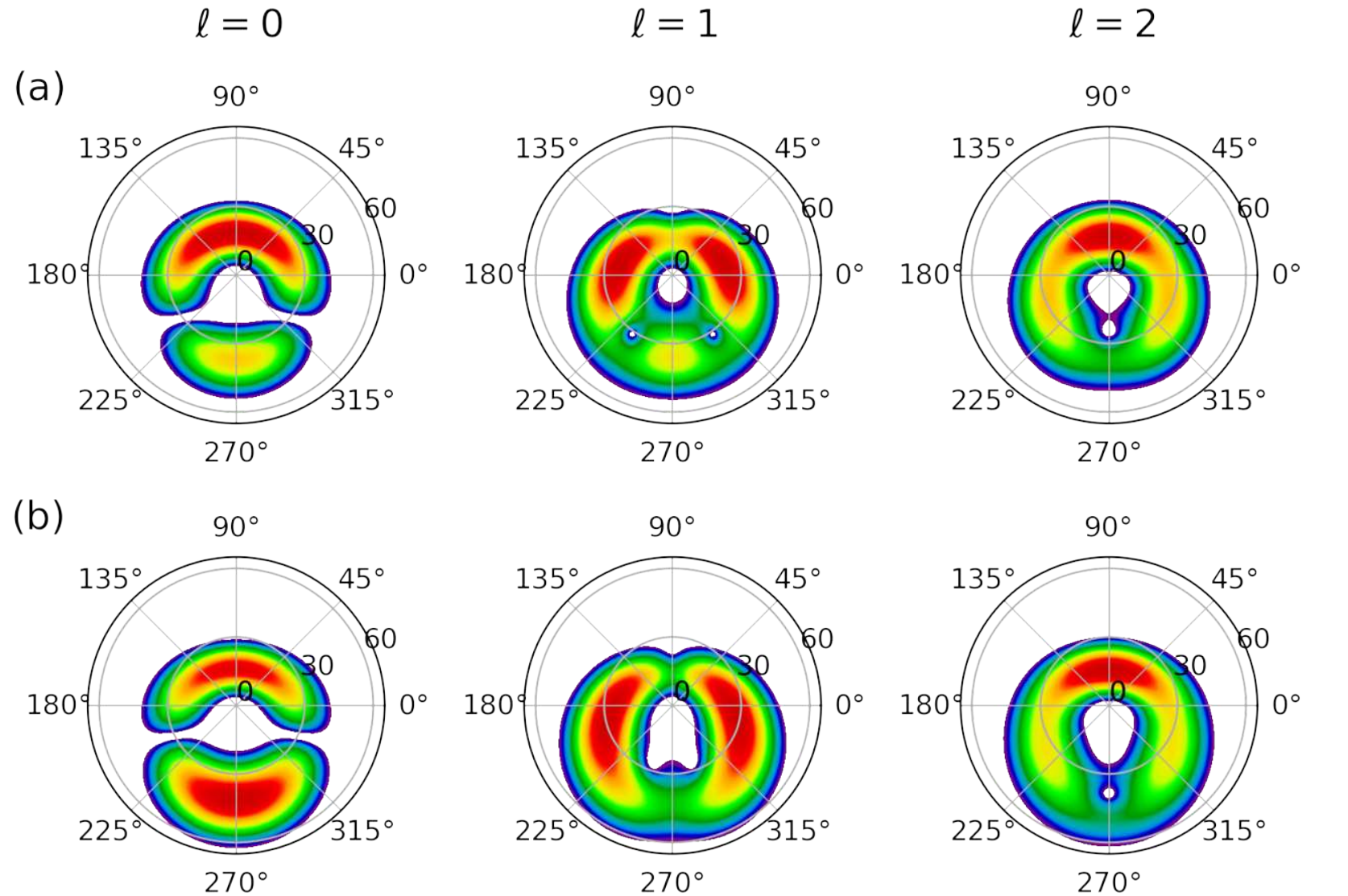}
\caption{Same as Fig.~\ref{fig2} except the liberated photoelectrons are streaked 
by the left-handed circularly polarised IR pulse. The spectra correspond to 
(a) $\theta_{k} = 30^{\circ}$ and (b) $\theta_{k} = 45^{\circ}$.}
\label{fig3}
\end{figure}

The presence of the three intensity maxima (and lobes) in the spectra for $l=1$ at $\theta_{k}=30^{\circ}$
could be understood as: 
The streaked part of Re($Y_{2}^{2}$) did not cancel  completely by the contribution  of
Re($Y_{2}^{0}$); and Im($Y_{2}^{2}$) contributes in the diagonal directions with respect to the cartesian axes whereas Im($Y_{2}^{0}$) is zero. 
For $l = 2$, the presence of the maxima and minima in intensity at $\theta_{k}=30^{\circ}$ is decided dominantly  by the 
first ionisation path, which is governed by $Y_{3}^{3}$. 
As the value of $\theta_{k}$ increases, other ionisation paths ($Y_{2}^{2}$ for $l=1$; $Y_{1}^{1}$ and $Y_{3}^{1}$ for $l=2$) contribute significantly, which results in the   
suppression of one of the three lobes and decides the 
angular modulation of the streaking spectra [middle and right columns in Fig. \ref{fig2}]. 
For $l=2$, a major distinction in the streaking spectra is observed at $\theta_{k}=60^{\circ}$. 
The reason lies in the angular structure of $Y_{3}^{1}$, which governs the third ionisation path. 
For different values of $\theta_{k}$, different lobes of $Y_{3}^{1}$ contribute with a threshold at 
approximately $\theta_{k}=50^{\circ}$, i.e., the contribution of $Y_{3}^{1}$ reverses when $
\theta_{k}$ crosses $50^{\circ}$. Thus giving similar spectra for $\theta_{k}=30^{\circ}$ and $\theta_{k}=45^{\circ}$; and for $\theta_{k}=60^{\circ}$ and $\theta_{k}=90^{\circ}$ (not shown here). The distinction and uniqueness of the streaking spectra, i.e., the angular 
modulation (maxima and minima)  in the intensity profile along with the strength of the minima, prove useful to discern the units of OAM encoded in the twisted pulse.

\begin{figure}[t!]
\includegraphics[width=8.5 cm]{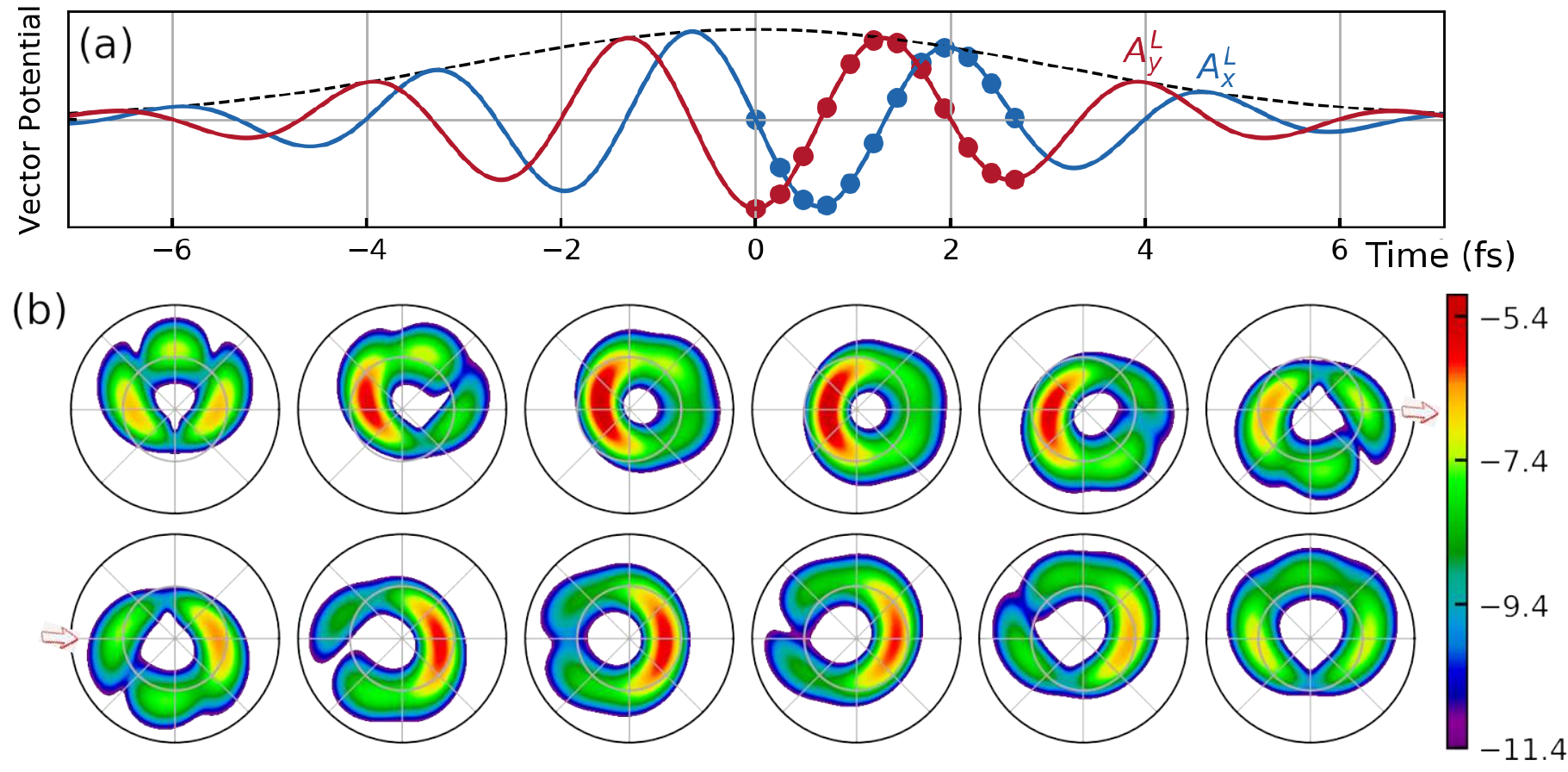}
\caption{(a) Vector potential of the right-handed circularly polarised   IR streaking pulse. The $x$- and $y$-components 
are shown in blue and red colours, respectively. (b) For $l = 1$, the attosecond streaking spectra observed at $\theta_{k}=30^{\circ}$ corresponding to the various time-delays between the IR and twisted pulse. The twisted pulse is delayed in steps of $10$ a.u. for one complete cycle of the IR pulse ($2.67$ fs or $110$ a.u.).}
\label{fig_delay2}
\end{figure}

Furthermore, instead of  right-handed circularly polarised IR pulse, 
left-handed circularly polarised IR pulse is used  for streaking  
to explore any possibility of the circular dichroism. 
Figure~\ref{fig3} presents the streaking spectra corresponding to left-handed circularly polarised IR streaking pulse for $\theta_{k} = 30^{\circ}$ and $45^{\circ}$. 
In this case, the vector potential of the IR pulse is along $+y$ direction, which 
results in streaking of photoelectrons along $-y$ direction. Therefore, 
all the spectra are streaked along the negative direction in comparison to the 
case when right-handed circularly polarised IR pulse is used to streak the 
photoelectrons.  It is established from Figs.~\ref{fig2} and ~\ref{fig3} that the 
streaking spectra are distinct and unique for each values of the OAM encoded 
in the  twisted  pulse. 

Till now we have discussed AAS when the time-delay between both the pulses is zero. 
To explore how the streaking spectrum changes as the time-delay between both the pulses 
is varied, several streaking spectra, along with the $x$- and $y$-components of the 
right-handed  
IR pulse, for $l = 1$ at $\theta_{k} = 30^{\circ}$ is presented in Fig.~\ref{fig_delay2}. 
The twisted pulse is delayed in steps of $10$ a.u. for a complete cycle of the IR pulse.
As the time-delay is increased, the vector potential of the IR pulse rotates clockwise starting from  $-y$ direction. 
So, the photoelectrons are streaked along $+y$ direction initially 
and later follows the $\textbf{A}_{L}(t)$ accordingly. Also, the intensity of the spectra follow 
the envelope of the IR pulse.  
It is evident  from the streaking spectra that the number of lobes are preserved as the time-delay is varied. 

In conclusion, 
we can summarise that the AAS spectra encode the 
signature and the amount of the OAM present in the twisted attosecond pulse. 
As a result in the modifications of the selection rules, several ionising 
paths by twisted pulse are possible, which results in angular modulations of the photoelectron spectra. 
By choosing specific observation directions of the photoelectrons, AAS technique is useful for linear, right as well as left circular polarisations of the unknown pulse. 
We believe that our proposed approach will be the core ingredient  in the light-matter interaction induced by twisted attosecond pulses. The present work paves the route  
to study helical light-matter interaction and 
twisted pulse mediated strong-field ionization via streak camera~\cite{kubel2017streak}.

G. D. acknowledges support from Science and Engineering Research Board (SERB) India 
(Project No. ECR/2017/001460) and Max-Planck India visiting fellowship. 


\end{document}